\documentclass[aps,prd,twocolumn,showpacs,preprintnumbers,amsmath,amssymb,nofootinbib,superscriptaddress]{revtex4}

\usepackage{graphicx}
\usepackage{dcolumn}
\usepackage{bm}
\usepackage{hyperref}

\usepackage{color}

\newcommand{\dd}{{\rm d}}

\begin{document}

\preprint{ICRR-Report-594-2011-11}
\preprint{YITP-11-85}

\title{Weak lensing of CMB by cosmic (super-)strings}

\author{Daisuke Yamauchi}
\email{yamauchi@icrr.u-tokyo.ac.jp}
\affiliation{
Institute for Cosmic Ray Research, The University of Tokyo,
Kashiwa 277-8582, Japan
}
\affiliation{
Yukawa Institute for Theoretical Physics, Kyoto University,
Kyoto 606-8502, Japan
}
\author{Keitaro Takahashi}
\email{keitaro@sci.kumamoto-u.ac.jp}
\affiliation{
Faculty of Science, Kumamoto University,
2-39-1, Kurokami, Kumamoto 860-8555, Japan
}
\affiliation{
Department of Physics and Astrophysics, Nagoya University,
Nagoya 494-8602, Japan
}
\author{Yuuiti Sendouda}
\email{sendouda at cc.hirosaki-u.ac.jp}
\affiliation{
Graduate School of Science and Technology, Hirosaki University,
Hirosaki, Aomori 036-8561, Japan
}
\affiliation{
APC, Universit\'e Paris 7,
75205 Paris, France
}
\author{Chul-Moon Yoo}
\email{yoo@yukawa.kyoto-u.ac.jp}
\affiliation{
Yukawa Institute for Theoretical Physics, Kyoto University,
Kyoto 606-8502, Japan
}

\date{\today}

\begin{abstract}
We study the effect of weak lensing by cosmic (super-)strings on the anisotropies of cosmic microwave background (CMB).
In developing a method to calculate the lensing convergence field due to strings, and thereby temperature and polarization angular power spectra of CMB, we clarify how the nature of strings, characterized by the intercommuting probability, can influence these CMB anisotropies.
Assuming that the power spectrum is dominated by Poisson-distributed string segments, we find that the convergence spectrum peaks at low multipoles and is mostly contributed from strings located at relatively low redshifts.
As the intercommuting probability decreases, the spectra of the convergence and hence the lensed temperature and polarizations are gained because the number density of string segments becomes larger.
An observationally important feature of the string-induced CMB polarizations is that the string-lensed spectra decay more slowly on small scales compared with primordial scalar perturbations from standard inflation.
\end{abstract}

\maketitle

\section{\label{sec:introduction}Introduction}

Although there has been no direct evidence for the detection of topological defects in the Universe, there are good theoretical reasons for believing their existence and reasonable prospects for the detection.
It is naturally expected that topological defects, appearing as solutions to the field equation in various models of particle physics, have formed during phase transitions in the early universe through spontaneous symmetry breakings~\cite{Kibble:1976sj,Hindmarsh:1994re,Perivolaropoulos:2005wa}.
Cosmic strings, as a remnant of unified theories, have been expected to offer a good opportunity to probe extremely high energy physics through their discovery~\cite{Jeannerot:2003qv}.

Recently, cosmic strings have attracted a renewed interest in the context of string cosmology since it was pointed out that a new type of cosmic strings, so-called cosmic superstrings, may be formed at the end of stringy inflation~\cite{Sarangi:2002yt,Jones:2003da,Copeland:2003bj,Dvali:2003zj,Kachru:2003sx} (for reviews, see \cite{Polchinski:2004ia,Polchinski:2004hb,Davis:2005dd,Copeland:2009ga,Sakellariadou:2009ev,Ringeval:2010ca,Majumdar:2005qc,Copeland:2011dx}).
Cosmic superstrings, stretched to macroscopic scales by the succeeding expansion of the Universe after their formation, could be also an observable remnant of the string theory as the most promising modern unified theory~\cite{Witten:1985fp}.

One of the observationally important features of cosmic superstrings compared to field-theoretic ones arises from the fact that, from the four-dimensional point of view, the process of intercommutation of two strings is probabilistic;
the intercommuting probability for field-theoretic strings $ P $ has been shown to be almost always unity~\cite{Shellard:1987bv,1988ComPh2:51M,Moriarty:1988fx,Achucarro:2006es,Achucarro:2010ub,Hashimoto:2005hi,Eto:2006db,Hanany:2005bc}, while the value of $ P $ can be significantly smaller for cosmic superstrings~\cite{Hanany:2005bc,Polchinski:1988cn,Jackson:2004zg}.
The intercommutation process provides an essential mechanism for a string network to lose its energy, letting the network approach to an attractor solution.
Therefore, observations associated with the global structure of the string network, which sensitively depends on the intercommuting probability, have a potential to give us important information about the nature of cosmic (super-)strings.

In particular, the imprint of cosmic strings on the cosmic microwave background (CMB) has been widely studied.
For example, strings induce CMB temperature anisotropies with an amplitude typically given by the dimensionless tension $G\mu$ through the so-called Gott-Kaiser-Stebbins (GKS) effect~\cite{Kaiser:1984iv,Gott:1984ef}.
Historically, cosmic strings were even seriously considered as seed of structure formation in the Universe~\cite{Vilenkin-Shellard,Vilenkin:1981iu}, though the observations of the acoustic oscillation in the CMB temperature power spectrum have excluded strings as a dominant source of the large-scale temperature anisotropy~\cite{Komatsu:2010fb}.
Using this fact, constraints on the tension for conventional field-theoretic strings have been derived~\cite{Bevis:2007gh,Bevis:2010gj,Battye:2010xz,Pogosian:2008am}.
Nevertheless, it has been argued that there are some possibilities to detect a signal of cosmic strings in CMB as we remind readers below.

First, the CMB temperature fluctuations can be dominated by those due to strings on small scales because there the primary fluctuations are damped out~\cite{Bevis:2010gj,Fraisse:2007nu,Pogosian:2008am,Yamauchi:2010ms,Hindmarsh:1993pu}.
Observations of the temperature angular power spectrum on small scales by, e.g., Atacama Cosmology Telescope (ACT)~\cite{Dunkley:2010ge,Fowler:2010cy} and South Pole Telescope (SPT)~\cite{Lueker:2009rx} could be useful for the search for cosmic strings.

Next, cosmic strings are themselves highly nonlinear objects that should induce non-Gaussian fluctuations.
Because the primordial fluctuations from standard inflation are expected to be very close to Gaussian, non-Gaussian features could serve as an effective tool to detect and probe cosmic strings.
The numerical simulations performed by Fraisse \textit{et al.}~\cite{Fraisse:2007nu}
have shown that the obtained one-point probability distribution function has a large deviation from the Gaussian distribution.
In particular, the one-point probability distribution function has negative skewness and a non-Gaussian tail.
These non-Gaussian statistics could be distinctive features from primary fluctuations and other secondary effects, and may enhance the observability of cosmic strings in future small-scale observations~\cite{Ringeval:2010ca,Takahashi:2008ui,Yamauchi:2010vy,Gangui:2001fr,Hindmarsh:2009qk,Hindmarsh:2009es}.

Finally, the CMB polarizations, especially the B-mode could be a useful probe of cosmic strings through the future observations such as PLANCK~\cite{:2006uk}, CMBPol~\cite{Baumann:2008aq}, ACTPol~\cite{arXiv:1006.5049}, SPTPol~\cite{2009AIPC.1185..511M}, and LiteBIRD~\cite{LiteBIRD}.
So far, at least two sources of the B-mode due to cosmic strings have been studied.
One is gravitational wave bursts from the small-scale structures on strings~\cite{Kawasaki:2010iy} and the others are vector and tensor perturbations produced by a string network itself;
recent numerical simulations~\cite{Bevis:2010gj,Bevis:2007qz,Pogosian:2007gi,Seljak:2006hi,Wyman:2005tu,Pogosian:2003mz} showed that the contribution from the network of conventional field-theoretic strings can dominate the B-mode spectrum over the angular range $150 < \ell <1000$.

In this paper, we investigate gravitational weak lensing due to cosmic (super-)strings as yet another source of the B and other modes.
This is to be distinguished from the integrated Sachs-Wolfe effect, which corresponds to the angular scale where the primary fluctuations are damped, namely, small angular scale~\cite{Hindmarsh:1993pu}.
The B polarizations are induced by lensing as the partial conversion of the primary E polarizations.
First, we calculate the spectrum of the convergence field due to strings using the segment formalism developed in our previous paper~\cite{Yamauchi:2010ms}.
Then, we obtain the temperature and polarization spectra of the CMB, clarifying how they can vary according to the value of the intercommuting probability $P$.

The paper is organized as follows.
In Sec.~\ref{sec:weak lensing}, we first review the propagation of the cross-section of the congruence of null geodesic in a perturbed universe.
In Sec.~\ref{sec:application}, we discuss the application of the formalism to a cosmic string network.
After describing the basic equations governing a string network incorporating the intercommuting probability $P$, we derive angular power spectra of a convergence due to a string network.
Then, in Sec.~\ref{sec:lensed power spectra}
we show lensed CMB angular power spectra due to cosmic (super-)strings
when the gradient deflection is taken into account.
Finally, we summarize our results in Sec.~\ref{sec:summary}.

\section{\label{sec:weak lensing}Weak gravitational lensing}

We consider the deformation of the cross-section of a congruence of null geodesics under propagation in a perturbative universe~\cite{Bernardeau:2009bm,Seitz:1994xf,Kaiser:1996tp,Blandford:1991zz,Sasaki:1987ad,1961RSPSA.264..309S,Uzan:2000xv,deLaix:1996vc,deLaix:1997dj} (for reviews, see \cite{Lewis:2006fu,Perlick:2004tq,Bartelmann:1999yn,Sasaki:1993tu}).
The metric of the spacetime is assumed to have the form:
\begin{eqnarray}
&&\dd\tilde{s}^2=\tilde g_{\mu\nu}\dd x^{\mu}\dd x^{\nu}
=a^2(\eta)g_{\mu\nu}\dd x^{\mu}\dd x^{\nu}\,,
\end{eqnarray}
where $\tilde{g}_{\mu\nu}$ is the physical spacetime metric, $a(\eta)$ is a given function of time corresponding to the scale factor of a homogeneous and isotropic background, and $g_{\mu\nu}$ is the conformally related metric assumed to have the form
\begin{eqnarray}
&&g_{\mu\nu}=\bar g_{\mu\nu}+h_{\mu\nu}\,;\\
&&\bar g_{\mu\nu}\dd x^{\mu}\dd x^{\nu}
=-\dd\eta^{2}+\dd\chi^{2}+\chi^{2}\omega_{ab}\dd\theta^{a}\dd\theta^{b}\,,
\end{eqnarray}
where $h_{\mu\nu}$ is a small perturbation and $\omega_{ab}$ is the metric
on the unit sphere, i.e.\ $\omega_{ab}\dd\theta^{a}\dd\theta^{b}=\dd\theta^{2}+\sin^{2}\theta\dd\varphi^{2}$ in spherical coordinates.
In what follows, tensors defined on the physical spacetime and those on the unperturbed spacetime will be distinguished by the indication of a tilde ( $\tilde{}$ ) and bar ( $\bar{}$ ) as above.

We consider two geodesics $g_0:\bar{x}^{\mu}(v)$ and
$g:x^{\mu}(v)=\bar{x}^{\mu}(v)+\xi^{\mu}(v)$ with $v$ being
the affine parameter along the light ray.
We choose $\bar{x}^{\mu}(v)$ as a reference geodesic
and $\xi^{\mu}(v)$ is a deviation vector labeling 
the reference geodesic.
The affine parameter along $g_0$, $v$ increases 
with decreasing time and $v =0$ at $O$.
With these notations we can define the tangent vector along
the geodesic $g$ as $\tilde{k}^{\mu}\equiv\dd x^{\mu}/\dd v$.
This is a null vector satisfying the equations:
\begin{eqnarray}
\tilde g_{\mu\nu}\tilde k^{\mu}\tilde k^{\nu}=0\,,\ \ \ 
\frac{{\rm D}\tilde k^{\nu}}{\dd v}\equiv\tilde k^{\mu}\tilde\nabla_{\mu}\tilde k^{\nu}=0\,,
\label{eq:geodesic eq}
\end{eqnarray}
where $\tilde\nabla_{\mu}$ is the covariant derivative associated
to the physical metric $\tilde g_{\mu\nu}$.
It can be shown that $\tilde g_{\mu\nu}\tilde k^{\mu}\xi^{\nu}$ is constant along the geodesic and vanishes at $O$, so we have $\tilde g_{\mu\nu}\tilde k^{\mu}\xi^{\nu}=0$.
We denote by $\tilde u^{\mu}$ the observer's 4-velocity at $O$,
satisfying $\tilde g_{\mu\nu}\tilde{u}^{\mu}\tilde{u}^{\nu}=-1$.
The energy of a photon measured by the observer can be written as
$\tilde E\equiv -\tilde g_{\mu\nu}\tilde k^{\mu}\tilde u^{\nu}$.
We also introduce $\eta_{0}$ as the conformal time at $O$.

With these notations, we now describe the geodesic deviations.
It is useful to express a geodesic deviation equation in terms of
two-dimensional orthogonal spacelike basis along the light ray, 
$\tilde e^{\mu}_a$ with $a=1,2$, which satisfy
\begin{eqnarray}
\tilde g_{\mu\nu}\tilde e^{\mu}_a\tilde e^{\mu}_b=\omega_{ab}\,,\ \ 
\tilde g_{\mu\nu}\tilde k^{\mu}\tilde e^{\nu}_a
=\tilde g_{\mu\nu}\tilde u^{\mu}\tilde e^{\nu}_a=0\,.
\end{eqnarray}
We use the Latin indices starting from the letter $a$ ($a,b,\cdots$)
for the polarization basis.
The vectors $\{ \tilde k^{\mu}\,,\tilde u^{\mu}\,,\tilde e^{\mu}_a\}$ form a basis at $O$.
They can be parallel transported along the geodesic $ g $:
\begin{eqnarray}
\frac{{\rm D}\tilde{e}^{\mu}_a}{\dd v}=0\,,\ \ 
\frac{{\rm D}\tilde{u}^{\mu}}{\dd v}=0\,.
\end{eqnarray}

Since $\tilde g_{\mu\nu}\xi^{\mu}\tilde k^{\nu}=0$, the deviation vector along the geodesic 
can be expressed as
\begin{eqnarray}
\xi^{\mu}=\xi^{a}\tilde{e}_a^{\mu}+\xi^{0}\tilde{k}^{\mu}\,.
\end{eqnarray}
In terms of the projected deviation vector 
$\xi^{a}\equiv \tilde e^a_{\mu}\xi^{\mu}$,
the geodesic deviation equation in the conformal transformed spacetime 
can be written as~\cite{Seitz:1994xf,1961RSPSA.264..309S}
\begin{eqnarray}
\frac{\dd^{2}\xi^{a}}{\dd v^{2}}=\tilde{\cal T}^a{}_b\,\xi^{b}\,,
\label{eq:deviation eq}
\end{eqnarray}
where we have used the definition, 
$(\tilde\nabla_{\alpha}\tilde\nabla_{\beta}
-\tilde\nabla_{\beta}\tilde\nabla_{\alpha})\tilde k^{\mu}
=\tilde R^{\mu}{}_{\nu\alpha\beta}\tilde k^{\nu}$,
and $\tilde R_{\mu\rho\sigma\nu}$ is the Riemann tensor of the metric $\tilde g_{\mu\nu}$.
We denote by $\tilde{\cal T}^a{}_b$ the symmetric optical tidal matrix, which is defined as
\begin{eqnarray}
&&\tilde{\cal T}^a{}_b=-\tilde{R}_{\mu\rho\nu\sigma}
\tilde{k}^{\mu}\tilde{k}^{\nu}\tilde{e}^{\rho a}\tilde{e}^{\sigma}_b
\end{eqnarray}
Given the initial condition at the observer, $\xi_{a}|_O=0$, 
the solution of Eq.~\eqref{eq:deviation eq} can be rewritten in
the following form:
\begin{eqnarray}
\xi^{a}(v)=\tilde{\cal D}^a{}_b(v)
\left(\frac{1}{\tilde{E}_O}
\frac{\dd\xi^{b}}{\dd v}\biggl|_{O}\right)
\equiv\tilde{\cal D}^a{}_b(v)\,\dd\theta^{b}_{\rm I}\,,
\label{eq:Jacobi map def}
\end{eqnarray}
where $\tilde E_O=\tilde E|_{O}$, 
$\dd\theta^{a}_{\rm I}$ denotes the angular coordinates of the image at $O$,
$\tilde{\cal D}_a{}^b$ is the Jacobi map and it satisfies
\begin{eqnarray}
\frac{\dd^{2}}{\dd v^{2}}\tilde{\cal D}^a{}_b=\tilde{\cal T}^a{}_c\tilde{\cal D}^c{}_b\,,
\label{eq:D eq}
\end{eqnarray}
with the initial condition at the observer $O$:
\begin{eqnarray}
\tilde{\cal D}^a{}_b\bigl|_{O}=0\,,\ \ \ 
\frac{\dd}{\dd v}\tilde{\cal D}^a{}_b\Bigl|_{O}=\tilde{E}_O\,\delta^{a}{}_b\,.
\end{eqnarray}
The angular position of the source is related to the displacement vector as
$\dd\theta_{\rm S}^a\equiv\xi^{a}(v_{\rm S})/D_{\rm S}$,
where $D_{\rm S}=\sqrt{|\det\tilde{\cal D}^a{}_b(v_S)|}$ 
is the angular diameter distance at the source.
The expression Eq.~\eqref{eq:Jacobi map def} can be rewritten in the form
\begin{eqnarray}
\dd\theta_{\rm S}^a
=\frac{\tilde{\cal D}^a{}_b(v_{\rm S})}{D_{\rm S}}\dd\theta_{\rm I}^b
\equiv{\cal A}^a{}_b\dd\theta_{\rm I}^b\,.
\label{eq:amplification matrix def}
\end{eqnarray}
We have defined the amplification matrix ${\cal A}^a{}_b$ characterizing
the deformation of the shape of the background light.
\\

The conformal transformation 
$\tilde g_{\mu\nu}\mapsto g_{\mu\nu}$ maps
a null geodesic on $\tilde g_{\mu\nu}$ to 
a null geodesic on $g_{\mu\nu}$ with
the affine parameter transformed as 
$\dd v\mapsto \dd\lambda =a^{-2}\dd v$~\cite{Wald}.
Therefore, we can introduce the null vector 
$k^{\mu}$ tangent to the geodesic $g$ defined by
\begin{eqnarray}
k^{\mu}\equiv a^2\tilde k^{\mu}
=\frac{\dd x^{\mu}}{\dd\lambda}\,,
\end{eqnarray}
Then, Eqs.~\eqref{eq:geodesic eq} are rewritten as
$g_{\mu\nu}k^{\mu}k^{\nu}=0$ and $k^{\mu}\nabla_{\mu}k^{\nu}=0$.
For convenience, we also introduce the 4-velocity $u^{\mu}$ and
the polarization basis $e^{\mu}_a$ on the conformal transformed spacetime by
$u^{\mu}=a\,\tilde u^{\mu}$ and $e^{\mu}_a=a\,\tilde e^{\mu}_a$.
The energy of the photon
can be rewritten as $\tilde{E}=a^{-1}g_{\mu\nu}k^{\mu}u^{\nu}\equiv a^{-1}E$.

Using these transformations, the evolution equation for the Jacobi map \eqref{eq:D eq}
can be rewritten as~\cite{Bernardeau:2009bm,Lewis:2006fu}
\begin{eqnarray}
\frac{\dd^{2}}{\dd\lambda^{2}}{\cal D}^a{}_b={\cal T}^a{}_c{\cal D}^c{}_b\,,
\label{eq:conformal D eq}
\end{eqnarray}
with the boundary conditions: ${\cal D}^a{}_b|_O=0$, $\dd{\cal D}^a{}_b/\dd\lambda |_O=E_O\,\delta^{a}{}_b$,
where $E_O=E|_O$.
We have defined the Jacobi map ${\cal D}^a{}_b$ 
and the symmetric optical tidal matrix ${\cal T}^a{}_b$ in conformal transformed spacetime:
\begin{eqnarray}
&&{\cal D}^a{}_b=a^{-1}\tilde{\cal D}^a{}_b\,,\\
&&{\cal T}^a{}_b=-R_{\mu\rho\nu\sigma}k^{\mu}k^{\nu}e^{\rho a}e^{\sigma}_b\,,
\end{eqnarray}
where $R_{\mu\rho\nu\sigma}$ is the Riemann tensor of $g_{\mu\nu}$.
Hence, in the cosmological background, it is sufficient to perform 
the calculation without the Hubble expansion and reintroduce 
the scale factor at the end by rescaling the Jacobi map,
namely ${\cal D}^a{}_b\mapsto\tilde{\mathcal D}^a{}_b=a\,{\cal D}^a{}_b$.

To see the solution of the Jacobi map in the conformal transformed spacetime, 
we then expand ${\cal D}_{ab}$, and ${\cal T}_{ab}$ 
as ${\cal D}_{ab}=\bar{\cal D}_{ab}+\delta{\cal D}_{ab}$, 
and ${\cal T}_{ab}=\bar{\cal T}_{ab}+\delta{\cal T}_{ab}$.
Since $\bar{\cal T}_{ab}=0$ in the unperturbed spacetime,
the zeroth order Jacobi map trivially reduces to $\bar{\cal D}^a{}_b=E_O\lambda\,\delta^{a}{}_b$.
Plugging the zeroth order solution, 
we have the Jacobi map up to linear order~\cite{Bernardeau:2009bm}
\begin{eqnarray}
\delta{\cal D}^a{}_b(\lambda_{\rm S})
=E_O\int^{\lambda_{\rm S}}_0
\dd\lambda\left(\lambda_{\rm S} -\lambda\right)\lambda\,\delta{\cal T}^a{}_b(\lambda )\,,
\label{eq:linear order Jacobi map}
\end{eqnarray}
where $\delta{\cal T}^a{}_b=-\delta R_{\mu\rho\nu\sigma}\bar k^{\mu}\bar k^{\nu}\bar e^{\rho a}\bar e^{\nu}_b$.
Since we consider only a first order in metric perturbations, we can evaluate
the integral along the unperturbed path, namely Born approximation.

Reintroducing the scale factor, 
we then have the amplification matrix by using Eq.~\eqref{eq:amplification matrix def} as
\begin{eqnarray}
{\cal A}^a{}_b
=\delta^{a}{}_b+\int^{\lambda_{\rm S}}_0
\dd\lambda\frac{\left(\lambda_{\rm S} -\lambda\right)\lambda}{\lambda_{\rm S}}\,
\delta{\cal T}^a{}_b(\lambda )\,,
\label{eq:amplifcation matrix}
\end{eqnarray}
where we have used the zeroth order result, $D_S=a(\lambda_{S})E_O\lambda_{S}$.
In particular, the convergence $\kappa$ is defined in terms of the amplification matrix as
\begin{eqnarray}
&&\kappa\equiv 1-\frac{1}{2}{\rm Tr}\,{\cal A}^a{}_b
\,.\label{eq:convergence def}
\end{eqnarray}

\section{\label{sec:application}Cosmic strings}

\subsection{String network dynamics}

We briefly review an analytic model for the evolution of a cosmic superstring network.
We rely on the velocity-dependent one-scale model~\cite{Martins:2000cs,Martins:1996jp,Avgoustidis:2005nv,Martins:2003vd} and extend it taking the probabilistic nature of the intercommutation process into consideration~\cite{Takahashi:2008ui}.

In the original model, a string network is characterized by just two physical quantities: the correlation length $\xi$ and the root-mean-square (rms) velocity $v_{\rm rms}$.
The correlation length, which is supposed to characterize the interstring distance at the same time, is defined by
\begin{eqnarray}
\xi =\sqrt{\frac{\mu}{\rho_{\rm str}}}\,,
\end{eqnarray}
where $\rho_{\rm str}$ denotes the energy density of the strings incorporated in the network.

To describe the evolution of the network, it is convenient to work with the cosmic time $ t \propto \int a \mathrm d\eta $.
A Hubble-normalized variable $\gamma\equiv 1/H\xi$ is also introduced for convenience.

In addition to Hubble friction, we take into account the energy loss from the network due to loop formations assuming its rate is given by $\mathrm d\rho_\mathrm{str}/\mathrm dt = -\tilde{c}Pv_{\rm rms}\rho_{\rm str}/\xi$~\cite{Yamauchi:2010ms,Takahashi:2008ui,Yamauchi:2010vy,Avgoustidis:2005nv}, where $\tilde{c}\approx 0.23$ quantifies the efficiency of loop formation~\cite{Martins:2000cs,Martins:2003vd}.
In the case of (decelerated) power-law expansion, $ a \propto \eta^{\beta/(1-\beta)} \propto t^\beta $ with $ \beta $ being constant, we obtain the evolution equations for $\gamma$ and $v_{\rm rms}$ as, respectively,
\begin{eqnarray}
&&\frac{t}{\gamma}\frac{\dd\gamma}{\dd t}
=1-\beta -\frac{1}{2}\beta\tilde{c}Pv_{\rm rms}\gamma
-\beta v_{\rm rms}^2
\,,\label{eq:gamma eq}\\
&&\frac{\dd v_{\rm rms}}{\dd t}
=(1-v_{\rm rms}^2)H
\Bigl[k\gamma -2v_{\rm rms}\Bigr]\,, 
\label{eq:v_rms eq}
\end{eqnarray}
where $k$ is the momentum parameter and for it we use the following approximate form~\cite{Martins:2000cs}:
\begin{eqnarray}
k(v_{\rm rms})=\frac{2\sqrt{2}}{\pi}
\frac{1-8v_{\rm rms}^6}{1+8v_{\rm rms}^6}\,.
\end{eqnarray}

We assume that, by the time of the last scattering of the CMB photons, the network reaches the regime of \emph{scaling}, in which the correlation length $ \xi $ scales with the Hubble radius.
In the scaling regime $\gamma$ and $v_{\rm rms}$ stay constant in time.
Then in the matter-dominated era, we can algebraically solve Eqs.~\eqref{eq:gamma eq} and \eqref{eq:v_rms eq} for $\tilde{c}P \ll 1$ to obtain~\cite{Takahashi:2008ui}
\begin{eqnarray}
v_{\rm rms}^2\approx\frac{1}{2}\biggl[
1-\frac{\pi}{3\gamma}\biggr]\,,
\quad
\gamma\approx\sqrt{\frac{\pi\sqrt{2}}{3\tilde{c}P}}\,.
\label{eq:scaling sol}
\end{eqnarray}

Note that in our treatment the string energy density comes to obey an approximate proportionality $\rho_{\rm str}=\mu\,\gamma^{2}\,H^2\propto P^{-1}$.
This $P$ dependence is consistent with the numerical simulations in \cite{Sakellariadou:2004wq}, while \cite{Avgoustidis:2005nv} obtained a relatively weaker dependence, $\rho_\mathrm{str}\propto P^{-\alpha}$ with $ 0 < \alpha < 1$.
It is beyond the scope of the present paper but would be important to calculate more precisely the scaling values, $\gamma$ and $v_{\rm rms}$, to match numerical calculations.

\subsection{Convergence due to a string network}
\label{sec:convergence}

We assume that strings can be well approximated as Nambu-Goto strings and
the gravitational field of strings is sufficiently weak so we can solve
the linearized Einstein equations.
From Eqs.~\eqref{eq:amplifcation matrix}, \eqref{eq:convergence def}, 
the convergence $\kappa$ can be described by
the stress-energy tensor $T_{\mu\nu}$ as
\begin{eqnarray}
&&\kappa =1-\frac{1}{2}{\rm Tr}\,{\cal A}^a{}_b
\nonumber\\
&&\ \ 
=\frac{1}{2}\int^{\chi_{\rm S}}_0\dd\chi
\frac{(\chi_{\rm S}-\chi )\chi}{\chi_{\rm S}}R_{\mu\nu}\bar{K}^{\mu}\bar{K}^{\nu}
\biggl|_{\eta =\eta_0 -\chi}
\nonumber\\
&&\ \ 
=4\pi G\int^{\chi_{\rm S}}_0\dd\chi
\frac{(\chi_{\rm S}-\chi )\chi}{\chi_{\rm S}}T_{\mu\nu}\bar{K}^{\mu}\bar{K}^{\nu}
\biggl|_{\eta =\eta_0 -\chi}
\,,
\label{eq:reduced convergence}
\end{eqnarray}
where we have introduced $\bar{K}^{\mu}=\bar{k}^{\mu}/E_O$,
$\dd\chi =E_O\dd\lambda$.
In what follow, we consider a static observer and $\bar{K}^{\mu}=(1,\hat{\bm n})$.
We note that the conformal distance to the source is not directly observable and
we should express the convergence as a function of the redshift of the source~\cite{Sasaki:1987ad}.
Although one expects a small correction due to the perturbation of the redshift, 
the contribution of the redshift is neglected in this paper.

To see the contribution due to string segments, we should evaluate 
the stress-energy tensor of a string.
In the conformal transformed spacetime, a convenient choice of the gauge is the transverse gauge.
In this gauge, the stress-energy tensor can be described as~\cite{Vilenkin-Shellard}
\begin{eqnarray}
&&T^{\mu\nu}({\bm r},\eta )
=\int\dd\sigma
\tilde{T}^{\mu\nu}
\delta^{3}\left( {\bm r}-{\bm r}_{\rm L}(\sigma ,\eta )\right)
\,,
\label{eq:stress-energy}
\end{eqnarray}
where $\tilde{T}_{\mu\nu}$ is given as
\begin{eqnarray}
\tilde{T}^{\mu\nu}
=\mu\left(
\begin{array}{cc}
1 & \dot{r}_{\rm L}^k\\
\dot{r}_{\rm L}^l & \dot{r}_{\rm L}^k\dot{r}_{\rm L}^l-{r_{\rm L}^k}'{r_{\rm L}^l}'\\
\end{array}
\right)\,,
\label{eq:stress energy tilde}
\end{eqnarray}
and we have introduced the three-dimensional coordinate with the origin $O$ as ${\bm r}$
and the embedding function of the string position as ${\bm r}_{\rm L}={\bm r}_{\rm L}(\sigma ,\eta )$.
We use the Latin indices starting from $k$ with the range from $1$ to $3$
and the bold letters to label spacelike 3-vectors.
The dot and the prime denote the derivative with respect 
to $\eta$ and $\sigma$, respectively.

In Eq.~\eqref{eq:reduced convergence}, the energy-momentum tensor is properly evaluated along
the line of sight $ \hat{\bm n} $ in the following manner:
Once the comoving distance $ \chi $ measured from us is given, it fixes
the comoving vector $ {\bm r} = \chi \hat{\bm n} $ and
the conformal time $ \eta = \eta_0 - \chi $ simultaneously.
The Dirac delta function in the integral (27) then picks up the value
of the string energy-momentum tensor $ \tilde T^{\mu\nu} $ at
the possible intersection of the light ray and the string world sheet,
$ (\sigma,\eta ) = (\sigma_{\rm lc}(\chi ,\hat{\bm n}), \eta_0-\chi) $,
where
\begin{equation}
\chi \hat{\bm n}
= {\bm r}_{\rm L}(\sigma_{\rm lc}(\chi ,\hat{\bm n}), \eta_0-\chi )\,.
\end{equation}

In this paper, we use the formalism proposed in \cite{Yamauchi:2010ms}
for the angular power spectrum due to cosmic (super-)strings.
Since the observed sky map of the convergence due to string segments 
can appear as a superposition of those due to each segment,
the total contributions of the convergence can be decomposed 
into each contribution of each string segment.
In our treatment, we first introduce a segment index ``$(i)$''
to denote the contribution from each segment between last scattering surface (LSS) 
and the present.
For simplicity, we also assume that the $i$-th lens object 
is localized at $\chi =\chi_{\rm L}^{(i)}$,
namely, thin-lens approximation.
Under this approximation, the distance between the observer and the $i$ th lens object 
can be approximated as 
$|{\bm r}_{\rm L}^{(i)}(\sigma_{\rm lc}(\chi ,\hat{\bm n}),\eta_0 -\chi )|\approx\chi_{\rm L}^{(i)}$.
Hence, the stress-energy of the string network along the ray
can be well approximated as
\begin{eqnarray}
T_{\mu\nu}(\chi\hat{\bm n},\eta_{0}-\chi )
\bar{K}^{\mu}\bar{K}^{\nu}
\approx\sum_{(i)}\frac{1}{\chi_{\rm L}^{(i)}}
\Sigma^{(i)}(\hat{\bm n})\delta\left(\chi -\chi_{\rm L}^{(i)}\right)
\,,\label{eq:stress-energy decomposition}
\end{eqnarray}
where $\Sigma^{(i)}(\hat{\bm n})$ denotes integrated
energy-momentum of the $ i $ th string observed by the CMB photon~\cite{Uzan:2000xv}:
\begin{eqnarray}
&&\Sigma^{(i)}(\hat{\bm n})
\approx\int\dd s\mu_{\rm proj}^{(i)}(s)
\delta\left(\hat{\bm n}-\hat{\bm n}_{\rm L}^{(i)}(s)\right)\,,
\end{eqnarray}
where we have introduced the angular position of the $i$-th segment along the ray as 
$\hat{\bm n}_{\rm L}^{(i)}(s)\equiv{\bm r}_{\rm L}^{(i)}
(\chi_{\rm L}^{(i)}s,\eta_{0}-\chi_{\rm L}^{(i)})/\chi_{\rm L}^{(i)}$,
the world sheet angular coordinate 
as $\dd s=\dd\sigma_{\rm lc} /\chi_{\rm L}^{(i)}$ and $\mu_{\rm proj}^{(i)}(s)$
represents the projected energy density of the string segment,
which is defined as
\begin{eqnarray}
&&\mu_{\rm proj}^{(i)}(s)
=\mu\frac{(1+\hat{\bm n}_{\rm L}^{(i)}\cdot\dot{\bm r}_{\rm L}^{(i)})^2
-(\hat{\bm n}_{\rm L}^{(i)}\cdot{\bm r}_{\rm L}^{(i)'})^2}
{1+\hat{\bm n}_{\rm L}^{(i)}\cdot\dot{\bm r}_{\rm L}^{(i)}}
\,.\label{eq:mu_proj}
\end{eqnarray}
The value of $\mu_{\rm proj}(s)$ is normally of ${\cal O}(\mu )$ and 
the thin-lens approximation implies that the contributions from
$\hat{\bm n}_{\rm L}^{(i)}\cdot{\bm r}_{\rm L}^{(i)'}$
and $\hat{\bm n}_{\rm L}^{(i)}\cdot\dot{\bm r}_{\rm L}^{(i)}$
are small.
Then the projected energy density is well approximated as $\mu_{\rm proj}^{(i)}(s)\approx\mu$.

We note that the thin-lens approximation may not hold
in the case of strings since strings are extended and move 
with relativistic speed.
Let us estimate the range of parameters in which the thin-lens 
approximation is a valid approximation. 
Since a string segment can be treated as a lens object with a thickness $\sim\xi$\,,
the thin-lens conditions can be estimated as $\xi /D_{\rm L}^{(i)}\ll 1$ 
and $\xi /D_{\rm LS}^{(i)}\ll 1$\,,
where $D_{\rm L}^{(i)}=a(\eta_0 -\chi_{\rm L}^{(i)})\chi_{\rm L}^{(i)}$, 
$D_{\rm LS}^{(i)}=a(\eta_0 -\chi_{\rm S})(\chi_{\rm S}-\chi_{\rm L}^{(i)})$
are the angular diameter distances
from the $i$-th lens object to the observer and the source, respectively.
In the matter-dominated era, the conditions constrain the range of 
the redshift of the segment as $z^{(i)}\gg 0.4(\tilde{c}P/0.23)^{1/2}$\,.

To see the physical interpretation, we now switch to the physical spacetime.
Substituting Eq.~\eqref{eq:stress-energy decomposition} into Eq.~\eqref{eq:reduced convergence},
we have
\begin{eqnarray}
&&\kappa (\hat{\bm n})
\approx 4\pi G\mu\sum_{(i)}
\frac{D_{\rm LS}^{(i)}}{D_{\rm S}}\int\dd s\,\,
\delta\left(\hat{\bm n}-\hat{\bm n}_{\rm L}^{(i)}(s)\right)
\,,\label{eq:convergence}
\end{eqnarray}
where $D_{\rm S}=a(\eta_{0}-\chi_{\rm S})\chi_{\rm S}$ is the angular diameter distance
from the source to the observer.
We can evaluate the harmonic coefficient of the convergence, $\kappa_{\ell m}$, as
\begin{eqnarray}
&&\kappa_{\ell m}
=\int\dd\hat{\bm n}\, \kappa (\hat{\bm n})\, Y_{\ell m}^*(\hat{\bm n})
\nonumber\\
&&\ \ \ \ \ \ 
\approx\sum_{(i)}
4\pi G\mu\frac{D_{\rm LS}^{(i)}}{D_{\rm S}}\int\dd s
Y_{\ell m}^*\left(\hat{\bm n}_{\rm L}^{(i)}\right)
\nonumber\\
&&\ \ \ \ \ \ 
\equiv\sum_{(i)}\kappa_{\ell m}^{(i)}\,,
\end{eqnarray}
and then the angular power spectrum of the convergence can be written as
\begin{eqnarray}
&&C_{\ell}^{\kappa\kappa}
\equiv\frac{1}{2\ell +1}\sum_{m}
\Big\langle\bigl|\kappa_{\ell m}\bigl|^{2}\Big\rangle
\nonumber\\
&&\ \ \ \ \ \ 
=\frac{1}{2\ell +1}\sum_{m}
\biggl[
\Big\langle\sum_{(i)}\bigl|\kappa_{\ell m}^{(i)}\bigl|^{2}\Big\rangle
+\Big\langle\sum_{(i)\neq (j)}\kappa_{\ell m}^{(i)}\kappa_{\ell m}^{(j)*}\Big\rangle
\biggr]
\nonumber\\
&&\ \ \ \ \ \ 
\equiv C_{\ell}^{\kappa\kappa :1\text{seg}}
+C_{\ell}^{\kappa\kappa :2\text{seg}}
\,,
\end{eqnarray}
For the string segments, the ensemble average can be replaced
by averaging over the parameter space~\cite{Yamauchi:2010ms},
\begin{eqnarray}
\Big\langle\cdots\Big\rangle\rightarrow 
\prod_{i=1}^N\biggl[
\frac{1}{N}\int\dd z^{(i)}\frac{\dd V}{\dd z^{(i)}}
\int\dd{\bm \Theta}_{\rm L}^{(i)}\cdot\frac{\dd n}{\dd{\bm \Theta}_{\rm L}^{(i)}}
\biggr]\cdots\,,
\end{eqnarray}  
where $(\dd V/\dd z)\dd z$ is the differential comoving volume element
at redshift $z$, $(\dd n/\dd{\bm\Theta}_{\rm L})\cdot\dd{\bm \Theta}_{\rm L}$ 
is the comoving number density of 
string segments with the parameters in the range 
$[{\bm \Theta}_{\rm L},{\bm \Theta}_{\rm L}+\dd{\bm \Theta}_{\rm L}]$.
Note that the total number of the segments can be rewritten as
$N=\int\dd z(\dd V/\dd z)\int (\dd n/\dd{\bm \Theta}_{\rm L})\cdot\dd{\bm \Theta}_{\rm L}$.

Now the string segments are assumed to be 
distributed randomly between LSS and the present 
consistently with the string network model.
This implies that there is no correlation between
two different segments, 
$\langle\kappa^{(i)}_{\ell m}\kappa^{(j)*}_{\ell m}\rangle=0$
for $(i)\neq (j)$.
Then, we can assume that the 1-segment contribution 
$C^{\kappa\kappa :1\text{seg}}_{\ell}$
dominates the angular power spectrum.
We should note that there may be nonzero contribution 
from the segment-segment correlation $C_\ell^{\kappa\kappa :2\text{seg}}$ 
at large scale $\ell >156(\tilde cP/0.23)^{-1/2}$\, if we consider a more
general string network~\cite{Yamauchi:2010ms}.
However, the estimation of $C_\ell^{\kappa\kappa :2\text{seg}}$ is 
beyond the scope of the present paper and we will simply assume
it is negligible.
We hope to come back to this issue in a future publication~\cite{Yamauchi:in progress}.
Then, the power spectrum of the convergence can reduce to
\begin{eqnarray}
C_{\ell}^{\kappa\kappa}
\approx\int^{z_{\rm LSS}}_{z_{\rm min}}\dd z\frac{\dd V}{\dd z}
\int\dd{\bm\Theta}_{\rm L}\cdot\frac{\dd n}{\dd{\bm\Theta}_{\rm L}}
\ {\cal G}_{\ell}^{\kappa\kappa}({\bm\Theta}_{\rm L},z)\,,
\label{C_l^kappakappa}
\end{eqnarray}
where we have introduced the minimum redshift $z_{\rm min}$
such that the optical depth becomes unity, namely
$\tau\approx\gamma\ln (1+z_\mathrm{min})=1$, namely
$z_{\rm min}\approx 0.5(\tilde{c}P/0.23)^{1/2}$~\cite{Takahashi:2008ui}.
And ${\cal G}_{\ell}^{\kappa\kappa}$
represents the angular power spectrum due to each segment:
\begin{eqnarray}
&&{\cal G}_{\ell}^{\kappa\kappa}({\bm\Theta}_{\rm L},z)
=\frac{1}{2\ell +1}\sum_{m}\bigl|\kappa_{\ell m}^{(i)}({\bm\Theta}_{\rm L},z)\big|^{2}
\nonumber\\
&&\ 
\approx\pi\left( G\mu\frac{D_{\rm LS}(z)}{D_{\rm S}}\right)^{2}
\nonumber\\
&&\ \ \ \times\int\dd s_1\dd s_2
P_{\ell}\left(\hat{\bm n}_{\rm L}(s_1;{\bm\Theta}_{\rm L})
\cdot\hat{\bm n}_{\rm L}(s_2;{\bm\Theta}_{\rm L})\right)
\,,\label{eq:G_l^kappakappa}
\end{eqnarray}
where $P_{\ell}(\cos\theta )$ is the Legendre function of the first kind and 
we have used the relation, 
$\sum_{m}Y_{\ell m}(\hat{\bm n}_1)Y_{\ell m}(\hat{\bm n}_2)
=(2\ell +1)P_{\ell}(\hat{\bm n}_1\cdot\hat{\bm n}_2)/4\pi$.

Assuming that a segment is uniformly distributed on the sky, 
we can set 
$\hat{\bm n}_{\rm L}(s_1)\cdot\hat{\bm n}_{\rm L}(s_2)
=\cos [|{\bm r}_{\rm L}'|(s_1-s_2)]$, and
we find that the dependence of the direction of the string segment
disappears in Eq.~\eqref{eq:G_l^kappakappa}.
Hence, the average over the string configuration parameters
can be replaced by the number density of the string network,
namely 
$\int\dd{\bm\Theta}_{\rm L}\cdot (\dd n/\dd{\bm\Theta}_{\rm L})\rightarrow 
\xi^{-3}=H^3\gamma^3$.

In our calculation, we consider only a segment 
of a long string with length $\sim\xi$ at each scattering.
Therefore we take the range of integration over $s_{1,2}$
as $|{\bm r}_{\rm L}'|s_{1,2}\leq\xi /D_{\rm L}\equiv 1/\ell_{\rm co}$,
where $\ell_{\rm co}$ is the angular scale corresponding to the correlation 
length of the segment.
Then, ${\cal G}_{\ell}^{\kappa\kappa}$ in Eq.~\eqref{eq:G_l^kappakappa} reduces to
\begin{eqnarray}
&&{\cal G}_{\ell}^{\kappa\kappa}(z)
=2\pi\left( G\mu\frac{D_{\rm LS}(z)}{D_{\rm S}}\right)^{2}
\nonumber\\
&&\ \ \ \ 
\times\frac{1}{|{\bm r}_{\rm L}'|^2\ell_{\rm co}(z)\,\ell}
\int^{2\ell /\ell_{\rm co}(z)}_{-2\ell /\ell_{\rm co}(z)}\dd u
P_{\ell}(\cos (u/\ell ))\,.
\label{eq:G_ell^kappakappa 2}
\end{eqnarray}
Once we set $|\dot{\bm r}_{\rm L}|=v_{\rm rms}$ and the Universe
is assumed to be matter-dominated, we can calculate the angular power
spectrum of the convergence by using Eqs.~\eqref{C_l^kappakappa} 
and \eqref{eq:G_ell^kappakappa 2}.

\begin{figure}[tbp]
 \begin{center}
  \includegraphics[width=80mm]{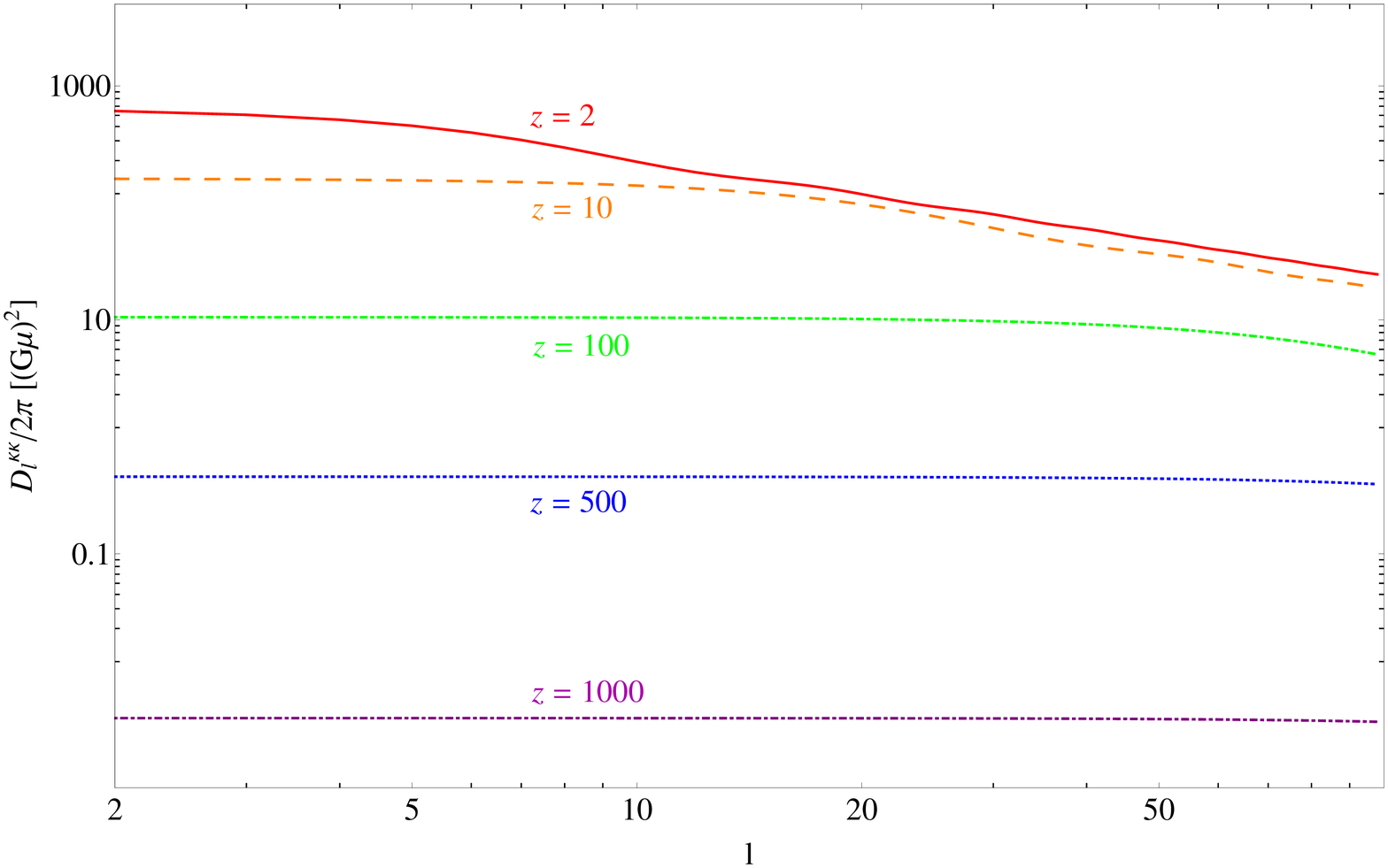}
 \end{center}
 \caption{\label{fig:Dlkappa_P1_z}
 The contributions to $C_{\ell}^{\kappa\kappa}$ from logarithmic intervals of $1+z$, 
$D_{\ell}^{\kappa\kappa}$, in the units of $(G\mu)^2$ for $P=1$.
 From top to bottom, $z=2$, $10$, $100$, $500$, and $1000$.} 
\end{figure}

\begin{figure}[tbp]
 \begin{center}
  \includegraphics[width=80mm]{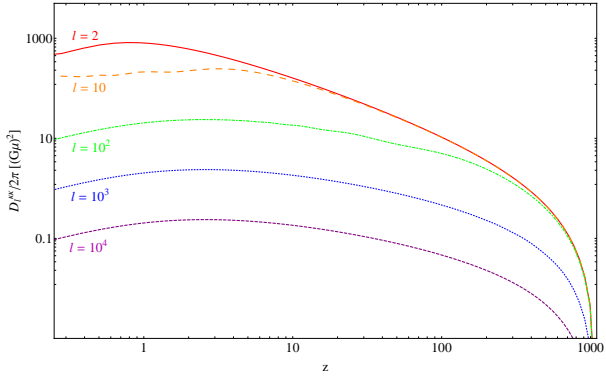}
 \end{center}
 \caption{\label{fig:Dlkappa_P1_l}
The same quantity as Fig.~\ref{fig:Dlkappa_P1_z} but as a function of $z$ for various values of $\ell$.
From top to bottom, $\ell =2$, $10$, $10^2$, $10^3$, and $10^4$.
 }
\end{figure}

Let us introduce the redshift distribution of $C_{\ell}^{\kappa\kappa}$\,:
\begin{eqnarray}
C_{\ell}^{\kappa\kappa}
=\int^{z_{\rm LSS}}_{z_\mathrm{min}}\frac{\dd z}{1+z}D^{\kappa\kappa}_{\ell}(z)\,.
\end{eqnarray}
In Figs.~\ref{fig:Dlkappa_P1_z} and \ref{fig:Dlkappa_P1_l},
$D_{\ell}^{\kappa\kappa}$ for $P=1$ is plotted as functions of $ z $ and $ \ell $, respectively.
We understand from these figures that the total amplitude mostly comes 
from the contributions of the segments at low redshifts, and the effect 
becomes more significant for lower multipoles.
This is because of the weight factor $(\chi_{\rm S}-\chi )\chi /\chi_{\rm S}$ 
in Eq.~\eqref{eq:reduced convergence}, which describes the fact that
the contribution from $\chi\approx \chi_{\rm S}/2$, namely $z\approx 2.5$, 
dominates the convergence field.

\begin{figure}[tbp]
 \begin{center}
  \includegraphics[width=80mm]{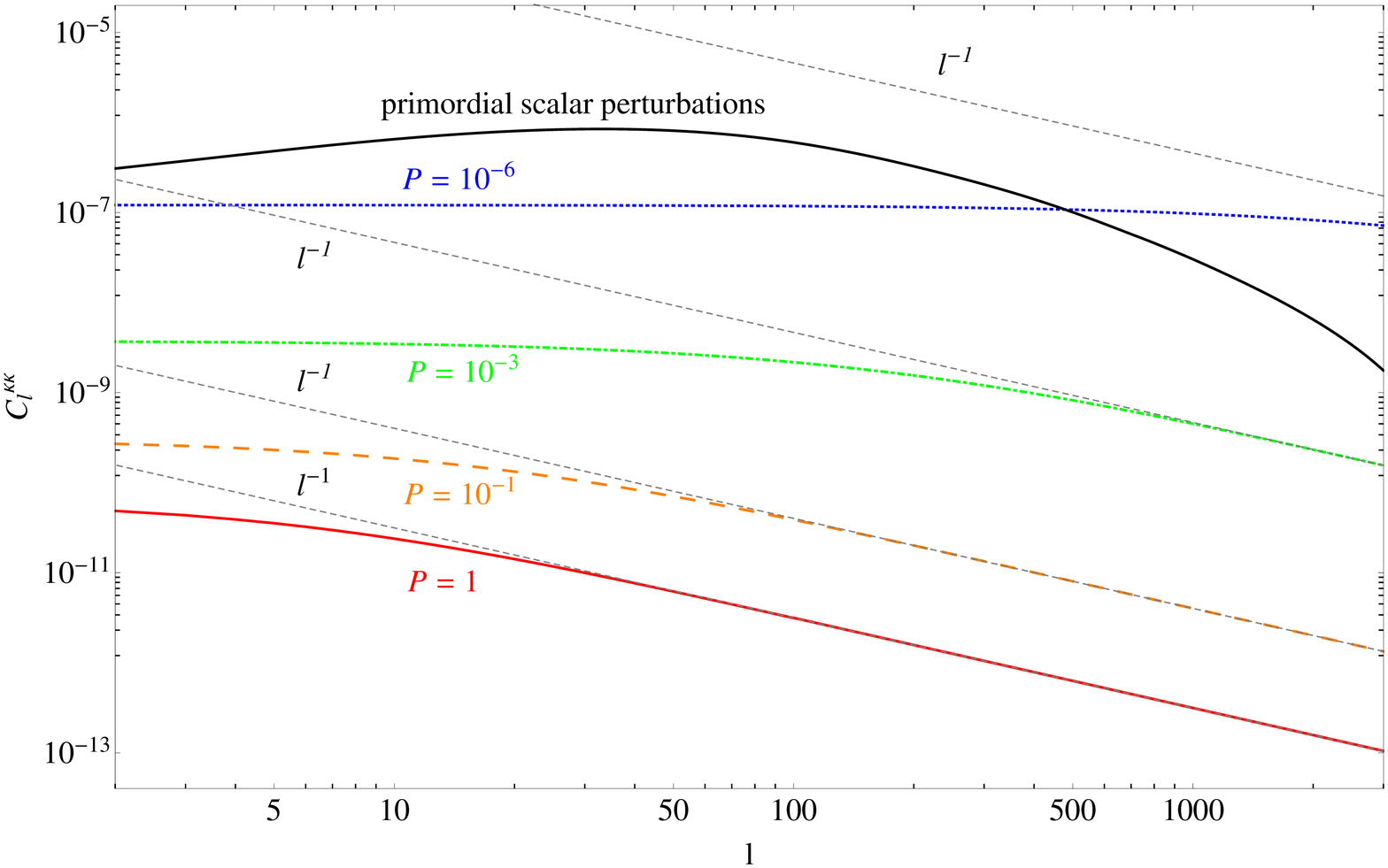}
 \end{center}
 \caption{\label{fig:Clkappa_P}
The all-sky total angular power spectrum $C_{\ell}^{\kappa\kappa}$ with $G\mu =2\times 10^{-7}$.
The curves are, from bottom to top, for $P=1$, $10^{-1}$, $10^{-3}$, and $10^{-6}$.
The gray dotted lines show the asymptotic behavior $\propto \ell^{-1}$ for large $ \ell $.
For comparison, the power spectrum of the convergence due to primordial scalar perturbations is shown by the black solid curve.
 } 
\end{figure}

In order to investigate the dependence on the intercommuting probability $P$\,, 
the total angular power spectra of the convergence $C_{\ell}^{\kappa\kappa}$\ 
with various values of $P$ are shown in Fig.~\ref{fig:Clkappa_P}.
The typical amplitudes of the lensing power spectrum at $\ell =10$ are 
$C_{\ell =10}^{\kappa\kappa}\approx 6.0\times 10^2(G\mu)^2$ for $P=1$, 
$4.6\times 10^3(G\mu)^2$ for $P=10^{-1}$,
$8.7\times 10^4(G\mu)^2$ for $P=10^{-3}$,
and $3.0\times 10^6(G\mu)^2$ for $P=10^{-6}$.

The spectrum generally shows the inverse power-law behavior, 
$\propto\ell^{-1}$, on small scales, while it has a plateau on large scales.
As $P$ decreases, the spectrum grows and the transition to the
inverse power-law occurs at smaller scale.
Actually, the amplitude on small scales $C_\ell^{\kappa\kappa}$
and transition multipole $\ell$ are found to be in proportion 
to $P^{-1}$ and $P^{-1/2}$\,, respectively.
These properties can be understood as follows:
Because $D_\ell^{\kappa\kappa}(z)\propto \xi (z)^{-3}\,{\cal G}_\ell^{\kappa\kappa}(z)$\,,
the dependence on $\ell$ is determined by ${\cal G}_\ell^{\kappa\kappa}(z)$\,.
Equation \eqref{eq:G_ell^kappakappa 2} implies that for small scales, $\ell >\ell_{\rm co}(z)$\,,
the integration over $u$ gives a constant value; therefore,
${\cal G}_\ell^{\kappa\kappa}(z)$ is in proportion to $\ell^{-1}$, 
while the dependence on $\ell$ becomes relatively weak for large scales, 
$\ell <\ell_{\rm co}(z)$\,.
Then integrating over $z$, we find that the transition of the spectrum
to the inverse power-law occurs at $\ell_{\rm co}(z_{\rm LSS})$, the $P$
dependence of it being $P^{-1/2}$\,.
One can also see that the dependence on $P$ of the amplitude can
be estimated as
$C_\ell^{\kappa\kappa}\propto\xi^{-3}\,\ell_{\rm co}^{-1}\propto P^{-1}$\,.

\section{\label{sec:lensed power spectra}Lensed power spectra from cosmic (super-)strings}

Lensing involves deflections that remap the temperature and polarization fields.
The lensing effect is conventionally expressed as the change of the direction vector by an angular gradient of the scalar lensing potential $\phi$:
\begin{eqnarray}
\hat{\bm n}\rightarrow \hat{\bm n}+\nabla_{\hat{\bm n}}\,\phi (\hat{\bm n})\,,
\label{eq:deflection angle}
\end{eqnarray}
where $\nabla_{\hat{\bm n}}$ represents the angular derivative, i.e., the covariant derivative on the sphere, and $\phi$ is defined in terms of the convergence as
\begin{eqnarray}
\kappa (\hat{\bm n})=\frac{1}{2}\nabla_{\hat{\bm n}}^{2}\,\phi (\hat{\bm n})\,.
\label{eq:lensing potential def}
\end{eqnarray}
This expression is valid for scalar perturbations, but topological defects also generate vector and tensor perturbations, which lead to a curly part of the deflection angle;
$\hat{\bm n}\rightarrow \hat{\bm n}+\nabla_{\hat{\bm n}}\,\phi 
+(*\nabla_{\hat{\bm n}})\,\varpi$,
where $*$ is the $90$-degree rotation
operator~\cite{Cooray:2005hm,Dodelson:2003bv,Cooray:2002mj,Hirata:2003ka,Stebbins:1996wx}.
The estimation of the curly component is beyond the scope of the present paper and we will simply assume it is negligible.
The effect of the curly component will be discussed in a forthcoming paper~\cite{Namikawa:2011cs,Yamauchi:2012bc}.
A brief derivation of the lensed angular power spectra is described in the Appendix, assuming the shape of the deflection takes the form of Eq.~\eqref{eq:deflection angle}.
Substituting the previously obtained power spectrum of the lensing potential
into Eqs.~\eqref{eq:lensed TT all sky}, \eqref{eq:lensed EE all sky}, \eqref{eq:lensed BB all sky}, and \eqref{eq:lensed TE all sky}, one obtains the all-sky CMB power spectra.

Although the string-induced convergence power spectrum peaks at low multipoles, higher multipole modes of CMB are also induced by lower multipole modes of lensing via convolution.
That is why we need to treat lensing
with all-sky formalism rather than flat-sky approximation.
In fact, numerical computations reveal that corrections to the flat-sky results amount to about $20\%$ even on small scales.

\begin{figure}[tbp]
 \begin{center}
  \includegraphics[width=80mm]{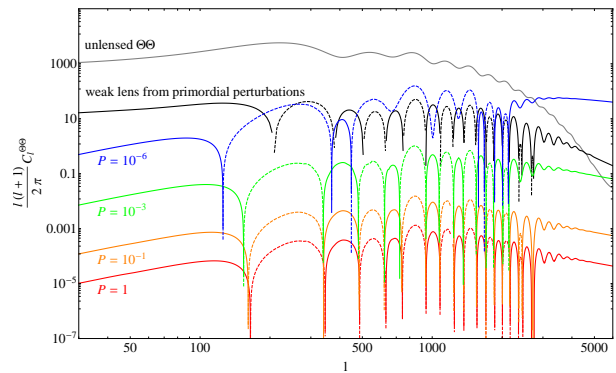}
 \end{center}
 \caption{\label{fig:ClTT_P}
The temperature angular power spectrum in the units of $\mu K^2$.
All the curves, except the unlensed primordial spectrum in gray, represent the difference of the lensed spectra from the unlensed one.
Parameters are, from bottom to top, $P=1$ (red), $10^{-1}$ (orange), $10^{-3}$ (green), and $10^{-6}$ (blue), with $G\mu =2\times 10^{-7}$.
The solid and dashed curves mean the positive and negative value, respectively.
The spectrum due to the primordial density perturbations is shown as the black curve for comparison.
}
\end{figure}
\begin{figure}[tbp]
 \begin{center}
  \includegraphics[width=80mm]{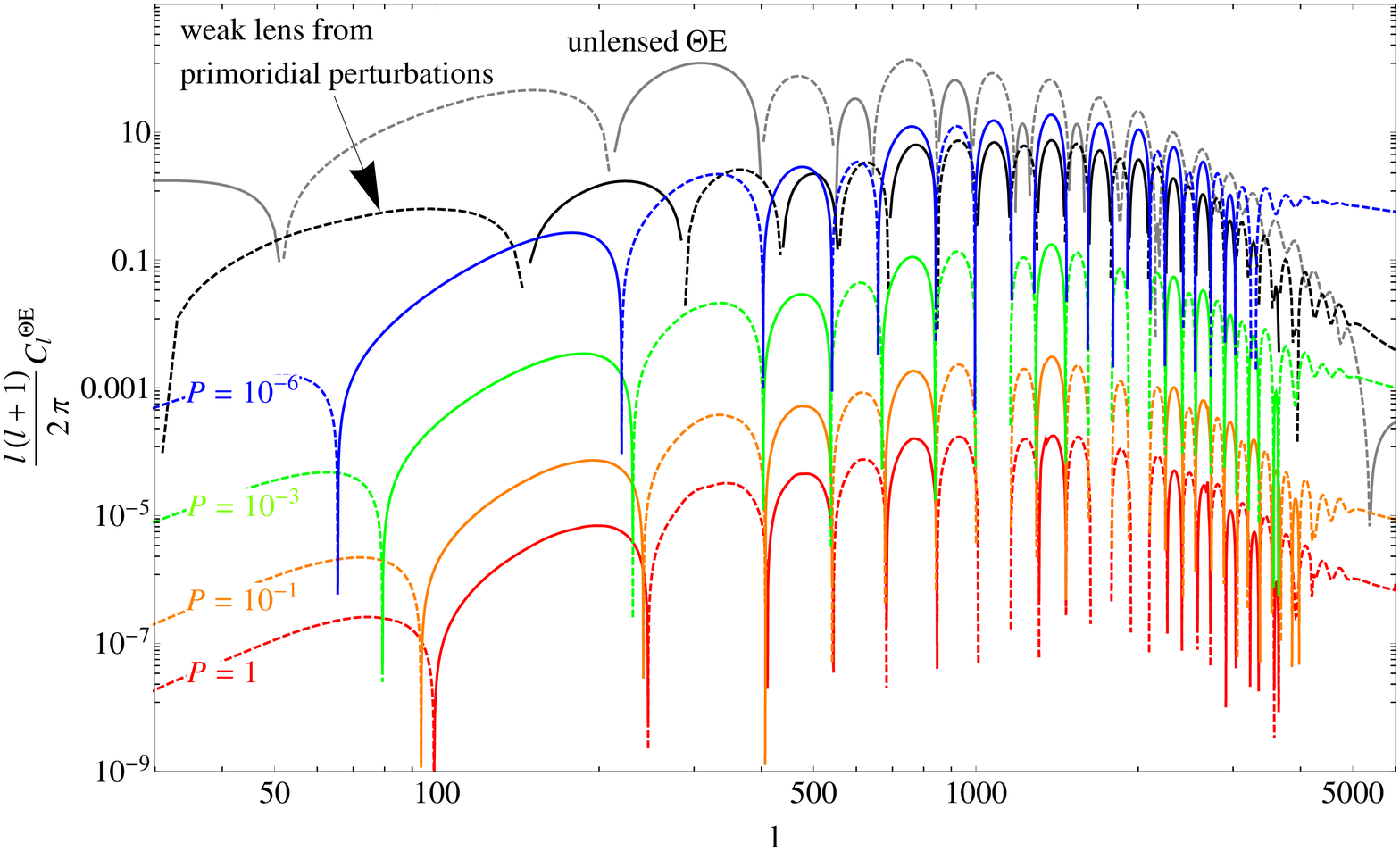}
 \end{center}
 \caption{\label{fig:ClTE_P}
The $\Theta$E polarization angular power spectrum in the units of $\mu K^2$.
The meaning of the curves and parameters are the same as Fig.~\ref{fig:ClTT_P}.
}
\end{figure}
\begin{figure}[tbp]
 \begin{center}
  \includegraphics[width=80mm]{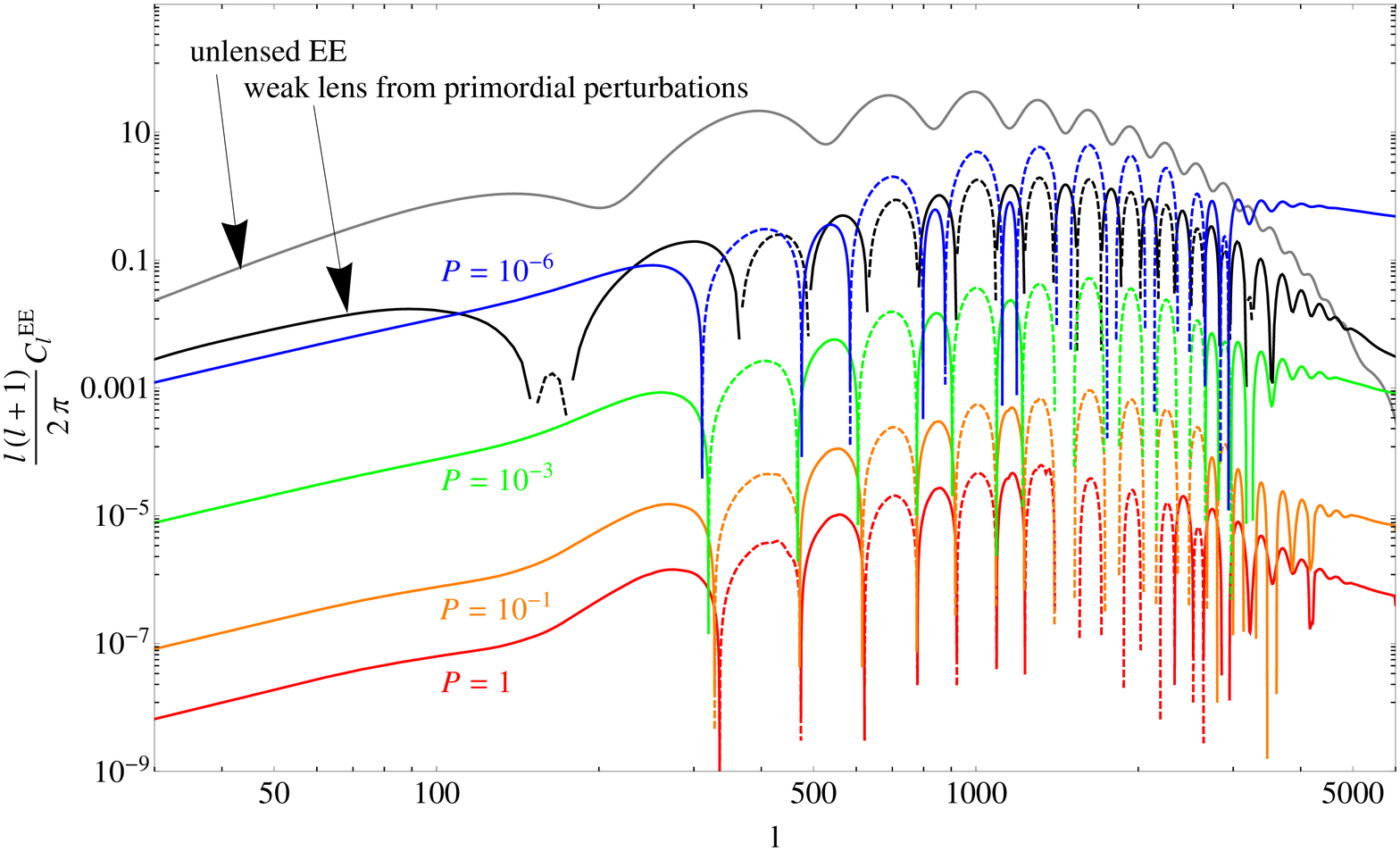}
 \end{center}
 \caption{\label{fig:ClEE_P}
The EE polarization angular power spectrum in the units of $\mu K^2$.
The meaning of the curves and parameters are the same as Fig.~\ref{fig:ClTT_P}.
 }
\end{figure}
\begin{figure}[tbp]
 \begin{center}
  \includegraphics[width=80mm]{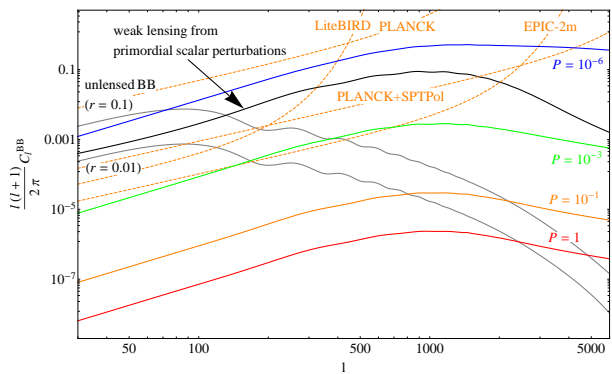}
 \end{center}
 \caption{\label{fig:ClBB_P}
The BB polarization angular power spectrum in the units of $\mu K^2$.
The curves are, from bottom to top, the difference between the lensed and unlensed spectra for $P=1$ (red), $10^{-1}$ (orange), $10^{-3}$ (green), and $10^{-6}$ (blue), with $G\mu =2\times 10^{-7}$.
The spectrum due to the primordial density perturbations is shown as the black curve for comparison.
The unlensed primordial spectra due to inflationary tensor perturbation with $r=0.1$ (not lensed by density perturbations) is shown by the gray solid line.
For comparison, the spectrum with $r=0.01$ is also shown by the gray solid line.
The dotted lines in orange are the sensitivity curve of PLANCK, EPIC-2m~\cite{Baumann:2008aq}, combined PLANCK+SPTPol~\cite{2009AIPC.1185..511M}, and LiteBIRD~\cite{LiteBIRD}.
 }
\end{figure}
Figures~\ref{fig:ClTT_P}, \ref{fig:ClTE_P}, \ref{fig:ClEE_P}, and \ref{fig:ClBB_P} respectively show the $\Theta\Theta$, $\Theta$E, EE and BB angular power spectra induced by the gravitational lensing due to cosmic (super-)strings.
For the string components, the normalization of the spectra is related to the string tension as $C_{\ell}^{{\rm X}{\rm X}'}\propto (G\mu )^2$, and we took $G\mu =2\times 10^{-7}$ for all the plots.
In these figures one can see that the amplitude of the spectra increases as $P$ decreases.
This is because the smaller the probability $ P $ is, the larger the amplitude of the power spectrum of the induced scalar lensing potential is.
For comparison, contributions from the inflationary density perturbations
are also shown.
The cosmological parameters are chosen to match the current CMB data~\cite{CAMB parameters}.

For $\Theta\Theta$, $\Theta$E, and EE, the differences between the lensed and unlensed spectra show the oscillatory behavior around zero, which is also seen in the case of the weak lensing due to primordial scalar perturbations~\cite{Hu:2000ee}.
For BB, in the large-scale limit the differences between the lensed and unlensed spectrum can be well approximated by white spectrum, namely $\tilde{C}_{\ell}^{\rm BB}-C_{\ell}^{\rm BB}={\rm constant}$.
On the other hand, an interesting feature observed in the small-scale limit is that the string-lensed BB spectra decay more slowly than that of the primordial scalar perturbations.
Actually, the spectra are approximately in proportion to the angular power spectrum of the convergence, $\ell^{2}(\tilde C_{\ell}^{\rm BB}-C_{\ell}^{\rm BB})\propto C_{\ell}^{\kappa\kappa}$~\cite{Lewis:2006fu}, and therefore we have that $ \ell^{2}(\tilde C_{\ell}^{\rm BB}-C_{\ell}^{\rm BB}) \propto \ell^{-1} $ (see Fig.~\ref{fig:Clkappa_P}).
This characteristic power-law behavior will help us distinguish cosmic (super-)strings from other secondary effects.
Even if the strings have only small tension, the string weak lensing could serve as the dominant source for the BB polarization on sufficiently small scales.

\begin{figure}[tbp]
 \begin{center}
  \includegraphics[width=80mm]{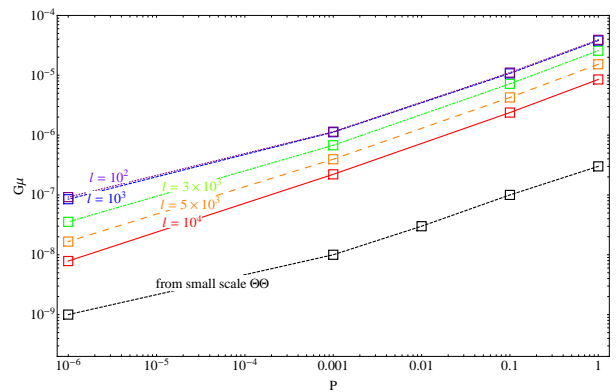}
 \end{center}
 \caption{\label{fig:Gmu-P_constraints}
 Constraints on $G\mu$ as a function of intercommuting probability $P$ obtained by requiring that the amplitude of the string-induced BB spectrum should be smaller than that of the primordial scalar perturbations at each $ \ell $\,.
For comparison, the upper bound obtained by considering the small-scale CMB temperature fluctuations from the GKS effect is shown in black solid curve~\cite{Yamauchi:2010ms}.}
\end{figure}
Because of how the amplitude depends on the intercommuting probability $ P $\,, the tension of strings $ G \mu $ is more tightly constrained for smaller values of $ P $\,.
Requiring the amplitude of the BB spectrum due to strings should be smaller than that due to the primordial perturbations at some $ \ell $ yields upper limits on $ G \mu $ as a function of the probability $ P $\,.
Figure~\ref{fig:Gmu-P_constraints} shows constraints on $ G \mu $ at $ \ell = 10^2 $, $ 10^3 $, $ 3 \times 10^3 $, $ 5 \times 10^3 $ and $ 10^4 $.
Tighter constraints on $G\mu$ can be obtained at larger $\ell$'s because the small-scale BB spectra decay more slowly compared with that due to the primordial perturbations, as we discussed above.

In Fig.~\ref{fig:Gmu-P_constraints} we have also plotted the upper bounds on the tension obtained by the comparison of the predicted small-scale temperature fluctuations due to the string GKS effect with observations~\cite{Yamauchi:2010ms}.
Although the constraints from the lensed B-mode are weaker than the upper bound on $G\mu$ from the GKS effect and the lensing contribution of the cosmic (super-)string network would be too small to be detected, the delensing of the CMB may substantially improve the effectiveness of the B-mode as a probe of cosmic strings in the future experiments~\cite{Seljak:2003pn,Sigurdson:2005cp}.

\section{\label{sec:summary}Summary}

In this paper, we gave calculations of cosmic (super-)string contributions to the temperature and polarization anisotropies of the CMB induced by gravitational weak lensing.
We developed a method to calculate the angular power spectrum of the convergence due to cosmic (super-)strings by using the segment formalism~\cite{Yamauchi:2010ms}.
Using our method, we clarified the dependence of the string-induced CMB polarizations on the intercommuting probability of strings explicitly.
Assuming that the power spectrum is dominated by Poisson-distributed string segments, we found that the contributions from segments located at low redshifts are dominant and low mutipole modes of lensing are essential even at high $\ell$ in CMB.
Therefore, we had to treat lensing effects due to a cosmic string network in an all-sky formalism.
The amplitude of the spectra increases as $P$ decreases, because the smaller the probability $ P $ is, the larger the amplitude of the power spectrum of the induced scalar lensing potential is.
An interesting feature observed in the small-scale limit is that the string-lensed spectra decay more slowly than that of the primordial scalar perturbations.

Let us argue an ambiguity in our calculation.
The correlation between two different segments has not been well understood either analytically or numerically.
Because the segment-segment correlation, if it exists, may not be negligible on large scale, 
the power spectrum of the scalar lensing potential may have additional modifications~\cite{Yamauchi:2010ms}.
It would be interesting to compare with numerical simulations and study
the large-angle behavior of power spectrum of the scalar lensing potential and 
the string-induced CMB polarizations.

It is also important to discuss lensing contributions to the CMB bispectrum.
Lensing events lead to deviations from Gaussianity because a lensed spectrum is a nonlinear function of two nearly Gaussian fields.
Equation~\eqref{eq:temp series expansion} implies the leading part of the bispectrum is in proportion to the cross-correlation between the temperature fluctuations and the lensing potential.
For cosmic (super-)strings, nonzero cross-correlation is expected to exist
between the integrated Sachs-Wolfe effect due to a straight string segment,
i.e.\ GKS effect~\cite{Kaiser:1984iv,Gott:1984ef},
and the gravitational potential included in the lensing potential $\phi$.
The work along this direction is in progress~\cite{Yamauchi:in progress}.

\acknowledgments

We thank M.~Hindmarsh, T.~Namikawa, M.~Sasaki, and A.~Taruya
for valuable comments and useful suggestions.
We also thank T.~Hiramatsu for useful discussion.
This work was supported in part by Monbukagaku-sho 
Grant-in-Aid for the Global COE programs, 
``The Next Generation of Physics, Spun from Universality 
and Emergence'' at Kyoto University and ``Quest for Fundamental Principles
in the Universe: from Particles to the Solar System and the Cosmos'' at
Nagoya University.
KT is supported by Grant-in-Aid for Young Scientists (B) No.~23740179.
DY was supported by Grant-in-Aid for JSPS Fellows No.~20-1117.
CY is supported by Grant-in-Aid for JSPS Fellows.
YS was supported by JSPS Postdoctoral Fellowships for Research Abroad.

\appendix

\section{\label{sec:lensed spectra}Lensed spectra}

We describe a simple derivation of the lensed angular power spectrum
using a series expansion in the deflection angle.
We begin by reviewing the calculations for the temperature and polarization 
lensed power spectrum, following \cite{Hu:2000ee}.
The CMB radiation field can be characterized by a $2\times 2$ intensity matrix $I_{ab}$.
The temperature fluctuation is given by $\Theta =(I_{11}+I_{22})/4$,
and the Stokes parameters $Q$ and $U$ are defined as $Q=(I_{11}-I_{22})/4$ and $U=I_{12}/2$.
In CMB observations, the temperature fluctuations on the sky, $\Theta (\hat{\bm n})$, can be decomposed into the multipole moments:
\begin{eqnarray}
&&\Theta (\hat{\bm n})=\sum_{\ell m}\Theta_{\ell m}Y_{\ell m}(\hat{\bm n})
\,.
\end{eqnarray}
Since the complex polarization fields, ${}_{\pm}X(\hat{\bm n})=Q(\hat{\bm n})\pm iU(\hat{\bm n})$, behave as spin-2 quantities, they can be decomposed in the harmonic coefficients for the spin-2 fields:
\begin{eqnarray}
&&{}_{\pm}X(\hat{\bm n})=\sum_{\ell m}{}_{\pm}X_{\ell m}
\,{}_{\pm 2}Y_{\ell m}(\hat{\bm n})
\,.
\end{eqnarray}
where ${}_{\pm 2}Y_{\ell m}$ is the spin-2 spherical harmonics on the unit sphere.
In particular, we can introduce the parity eigenstates:
\begin{eqnarray}
&&E_{\ell m}=\frac{1}{2}\Bigl({}_+X_{\ell m}+{}_-X_{\ell m}\Bigr)
\,,\\
&&B_{\ell m}=\frac{1}{2i}\Bigl({}_+X_{\ell m}-{}_-X_{\ell m}\Bigr)\,.
\end{eqnarray}

The lensing effects are conventionally expressed by the angular gradient of the lensing potential $\phi$:
\begin{eqnarray}
\hat{\bm n}\rightarrow \hat{\bm n}+\nabla_{\hat{\bm n}}\,\phi (\hat{\bm n})\,.
\end{eqnarray}
Through the gravitational lensing of the matter perturbation,
the temperature anisotropy of the CMB, 
$\Theta (\hat{\bm n})$, becomes
\begin{eqnarray}
&&\tilde{\Theta} (\hat{\bm n})
\equiv\Theta (\hat{\bm n}+\nabla_{\hat{\bm n}}\phi )
\nonumber\\
&&\ \ \ \ \ \ \ \ 
\approx\Theta (\hat{\bm n})
+\phi (\hat{\bm n})^{:a}\Theta (\hat{\bm n})_{:a}
\nonumber\\
&&\ \ \ \ \ \ \ \ \ \ \ 
+\frac{1}{2}\phi (\hat{\bm n})^{:a}
\phi (\hat{\bm n})^{:b}
\Theta (\hat{\bm n})_{:ab} +\cdots\,,
\label{eq:temp expansion}
\end{eqnarray}
where the colon ($\,:\,$) denotes the covariant derivative with respect to the metric on the sphere, $\omega_{ab}$\,.
Then, the harmonic coefficients are described by the unlensed
temperature fluctuations as~\cite{Hu:2000ee}
\begin{eqnarray}
&&\tilde{\Theta}_{LM}
=\Theta_{LM}
\nonumber\\
&&\ \ 
+\sum_{\ell ,\ell' ,m,m'}\phi_{\ell m}\Theta_{\ell' m'}\,
{}_0{\cal S}_{\ell\ell' L}^{\phi}(-1)^M
\left(
\begin{array}{ccc}
\ell &\ell' &L\\
m&m'&-M\\
\end{array}
\right)\nonumber\\
&&\ \ 
+{\cal O}\left(\phi^{2}\right)\,,
\label{eq:temp series expansion}
\end{eqnarray}
where  $\left(
\begin{array}{ccc}
\ell &\ell' &L\\
m&m'&-M\\
\end{array}
\right)$ denotes the Wigner $3j$ symbols and ${}_s{\cal S}^{\phi}$ is described as
\begin{eqnarray}
&&{}_{s}{\cal S}_{\ell\ell' L}^{\phi}=\frac{1}{2}
\biggl[\ell (\ell +1)+\ell' (\ell' +1)-L (L+1)\biggr]
\nonumber\\
&&\ \ 
\times
\sqrt{\frac{(2L+1)(2\ell +1)(2\ell' +1)}{4\pi}}
\left(
\begin{array}{ccc}
L &\ell &\ell'\\
s&0&-s\\
\end{array}
\right)
\,,\label{eq:S all sky def}
\end{eqnarray}

By following the same step as we did for the temperature fluctuations, the harmonic coefficients of the lensed complex Stokes parameters, ${}_{\pm}\tilde{X}_{\ell m}$, can be expand in the deflection angle as
\begin{eqnarray}
&&{}_{\pm}\tilde{X}_{LM}
={}_{\pm}X_{LM}
\nonumber\\
&&\ 
+\sum_{\ell ,\ell' ,m,m'}\phi_{\ell m}{}_{\pm}X_{\ell' m'}\,
{}_{\pm 2}{\cal S}_{\ell\ell' L}^{\phi}(-1)^M
\left(
\begin{array}{ccc}
\ell &\ell' &L\\
m&m'&-M\\
\end{array}
\right)\nonumber\\
&&\ 
+{\cal O}\left(\phi^{2}\right)\,,
\end{eqnarray}
where ${}_{\pm 2}{\cal S}^{\phi}$ was defined in Eq.~\eqref{eq:S all sky def}.

Because of a statistical anisotropy, the power spectrum and 
cross correlations of these quantities are defined by
\begin{eqnarray}
\big\langle x^*_{\ell m}x'_{\ell' m'}\big\rangle
=C_{\ell}^{xx'}\delta_{\ell\ell'}\delta_{mm'}\,.
\end{eqnarray}
where $\langle\cdots\rangle$ represents the ensemble average over the sky, 
$x$ and $x'$ can take on the values $\Theta$ ,$E$, $B$, and $\phi$.
Since the convergence field $\kappa$ is directly related to the lensing potential $\phi$ through 
Eq.~\eqref{eq:lensing potential def}, the power spectrum for the lensing potential $C_{\ell}^{\phi\phi}$ 
can be written in terms of the power spectrum for the convergence $C_{\ell}^{\kappa\kappa}$ as
\begin{eqnarray}
C_{\ell}^{\phi\phi}=\frac{4}{\ell^{2}(\ell +1)^2}C_{\ell}^{\kappa\kappa}\,.
\end{eqnarray}
Recalling ${}_{\pm}X_{\ell m}=E_{\ell m}\pm iB_{\ell m}$,
we have the lensed power spectra in terms of the power spectra
of the unlensed fields, $C_{\ell}^{\Theta\Theta}$, $C_{\ell}^{\Theta {\rm E}}$,
$C_{\ell}^{\rm EE}$ and $C_{\ell}^{\rm BB}$, as~\cite{Hu:2000ee}
\begin{eqnarray}
&&\tilde{C}_{\ell}^{\Theta\Theta}
=\Bigl\{ 1-\ell (\ell +1){\cal R}\Bigr\} C_{\ell}^{\Theta\Theta}
\nonumber\\
&&\ \ \ \ \ \ \ \ \ \ 
+\sum_{\ell_{1}\ell_{2}}\frac{({}_0{\cal S}_{\ell_{1}\ell_{2}\ell}^{\phi})^2}{2\ell +1}
C_{\ell_{1}}^{\phi\phi}C_{\ell_{2}}^{\Theta\Theta}
\,,\label{eq:lensed TT all sky}
\\
&&\tilde{C}_{\ell}^{\rm EE}
=\Bigl\{ 1-(\ell^{2}+\ell -4){\cal R}\Bigr\}
C_{\ell}^{\rm EE}
+\sum_{\ell_{1}\ell_{2}}
\frac{({}_{2}{\cal S}_{\ell_{1}\ell_{2}\ell}^{\phi})^2}{2(2\ell +1)}
C_{\ell_{1}}^{\phi\phi}
\nonumber\\
&&\ \ \ \times\biggl[
\left( C_{\ell_{2}}^{\rm EE}+C_{\ell_{2}}^{\rm BB}\right)
+(-1)^{\ell +\ell_{1}+\ell_{2}}
\left( C_{\ell_{2}}^{\rm EE}-C_{\ell_{2}}^{\rm BB}\right)\biggr]
\,,\label{eq:lensed EE all sky}\\
&&\tilde{C}_{\ell}^{\rm BB}
=\Bigl\{ 1-(\ell^{2}+\ell -4){\cal R}\Bigr\}
C_{\ell}^{\rm BB}
+\sum_{\ell_{1}\ell_{2}}
\frac{({}_{2}{\cal S}_{\ell_{1}\ell_{2}\ell}^{\phi})^2}{2(2\ell +1)}
C_{\ell_{1}}^{\phi\phi}
\nonumber\\
&&\ \ \ \times\biggl[
\left( C_{\ell_{2}}^{\rm EE}+C_{\ell_{2}}^{\rm BB}\right)
-(-1)^{\ell +\ell_{1}+\ell_{2}}
\left( C_{\ell_{2}}^{\rm EE}-C_{\ell_{2}}^{\rm BB}\right)\biggr]
\,,\label{eq:lensed BB all sky}\\
&&\tilde{C}_{\ell}^{\Theta {\rm E}}
=\Bigl\{ 1-(\ell^{2}+\ell -2){\cal R}\Bigr\}
C_{\ell}^{\Theta {\rm }E}
\nonumber\\
&&\ \ \ \ \ \ \ \ \ \ 
+\sum_{\ell_{1}\ell_{2}}
\frac{({}_0{\cal S}_{\ell_{1}\ell_{2}\ell}^{\phi})
({}_{2}{\cal S}_{\ell_{1}\ell_{2}\ell}^{\phi})}{2\ell +1}
C_{\ell_{1}}^{\phi\phi}C_{\ell_{2}}^{\Theta {\rm E}}
\label{eq:lensed TE all sky}\,,
\end{eqnarray}
where 
\begin{eqnarray}
&&{\cal R}
=\sum_{\ell_{1}}\ell_{1}(\ell_{1}+1)
\frac{2\ell_{1}+1}{8\pi}C_{\ell_{1}}^{\phi\phi}
\,.\label{eq:R all sky def}
\end{eqnarray}
The gravitational lensing contributes to EE and BB spectrum 
mainly through partial conversion of the EE spectrum.
Given the power spectra of the unlensed fields and lensing potential,
one can calculate the lensed spectra.



\begin{thebibliography}{}
\bibitem{CAMB parameters}
	The cosmological parameters we have used are $h=0.7$, $\Omega_{\rm b}h^2=0.0226$, 
$\Omega_{\rm c}h^2=0.114$, $\Delta_{\cal R}^2=2.46\times 10^{-9}$, $n_{\rm s}=0.96$ 
and $r=0.1$, with zero running of $n_{\rm s}$ and spatial flatness assumed.
\end{thebibliography}
\end{document}